\documentclass[leqno,12pt]{article}
\usepackage{amsmath,amssymb,amsthm,amsfonts,amscd}
\usepackage{hyperref}
\usepackage{multirow}
\usepackage[numeric]{amsrefs}
\input colordvi
\usepackage{graphicx}
\newcommand{\secref}[1]{Section~\ref{#1}}

\numberwithin{equation}{section}

\renewcommand\d{\delta}

\renewcommand\L{\Lambda}

\newcommand\f{\frac}
\newcommand\smallf[2]{{\textstyle{\frac{#1}{#2}}}}

\newcommand{\Z}{{\mathbb{Z}}}
\newcommand{\R}{{\mathbb{R}}}

\renewcommand\({\left(}
\renewcommand\){\right)}

\newcommand{\gobble}[1]{}
  \newcommand{\rangeref}[2]{%
    \ref{#1}--\afterassignment\gobble\fam 0\ref{#2}%
  }

\makeatletter
\def\imod#1{\allowbreak\mkern5mu({\operator@font mod}\,#1)}
\makeatother
\begin{document}

\title{Coppersmith's lattices and   ``focus groups'':~an   attack on small-exponent RSA}

\author{Stephen D. Miller\thanks{Supported by NSF grants CNS-1526333 and CNS-1815562. },
%\\
%Rutgers University\\
%\tt{miller@math.rutgers.edu}\\ \\
Bhargav Narayanan\thanks{Supported by NSF grant DMS-1800521.},  \\ %\\%University of Cambridge \\
\tt{\{miller,narayanan\}@math.rutgers.edu  }\\  and
Ramarathnam Venkatesan\\
%(Microsoft Research India)\\
\tt{venkie@microsoft.com}
}

\maketitle

\begin{abstract}  We present a principled technique for reducing the lattice and matrix size in some applications of Coppersmith's lattice method for finding roots of modular polynomial equations.  It relies on
extrapolating patterns from the  actual behavior of Coppersmith's attack for smaller parameter sizes, which can be thought of as ``focus group'' testing.  When applied to the small-exponent RSA problem, our technique reduces lattice dimensions and consequently  running times, and hence can be applied to a wider range of exponents.  Moreover, in many difficult examples our attack is not only faster but also more successful in recovering the RSA secret key. We include a discussion of subtleties concerning whether or not existing metrics (such as enabling condition bounds) are decisive in predicting the true efficacy of attacks based on Coppersmith's method.  Finally, indications are given which suggest certain lattice basis reduction algorithms (such as Nguyen-Stehl\'e's L2) may be particularly well-suited for Coppersmith's method.

\end{abstract}

\section{Introduction}\label{sec:intro}

Ever since Shamir's devastating attack on the Knapsack cryptosystem \cite{shamir}, lattice basis reduction algorithms such as LLL \cite{LLL} have found surprising success against cryptosystems that {\em a priori} have nothing to do with lattices.  A fundamental example is the \textsc{RSA} cryptosystem \cite{RSA}, whose public key consists of an integer $n=pq$ (where $p$ and $q$ are large secret primes of comparable size) and an encryption exponent $e$.  In situations where some extra information about the public key is known (e.g., certain bits of $p$ or $q$), it is sometimes possible to use the lattice basis reduction method of Coppersmith \cite{coppersmith1} to discover the factorization of $n$.

One notable such situation is when the secret decryption exponent $d$ is {\em small},
\begin{equation}\label{smallexp}
  d \ \ = \ \ n^\delta \ , \ \ \delta\,<\,\smallf 12
\end{equation}
(see \secref{sec:overview} for more background on RSA).  Wiener \cite{wiener} showed that continued fractions  expose $d$ when $\delta<1/4$, essentially instantaneously.
Continued fraction approximations can be thought of as the simplest example of lattice basis reduction, namely for 2-dimensional lattices.  Boneh-Durfee \cite{BD} apply  Coppersmith's method with higher dimensional lattices to give an attack for
\begin{equation}\label{bdexp}
  \delta \ \ < \ \ 1 \ - \ 2^{-1/2} \ \ \approx  \ \ .292\,.
\end{equation}
More precisely,
 they prove that LLL's output on a particular lattice   produces enough information to factor $n$ (subject to an algebraic independence assumption).  It is an important open problem to improve\footnote{This raises the question of   what  precisely constitutes an improvement over Boneh-Durfee's .292 result \cite{BD}.  Merely replacing .292 by a larger constant  alone is  insufficient, since algebraic dependence may creep for smaller $\delta$.  Indeed, Bauer \cite{bauerthesis} gives an attack
 in which the right-hand side of (\ref{bdexp}) can be replaced by .34, but which suffers from a failure of algebraic independence at a much earlier point.  On the other hand, it should be pointed out that   we don't even know at present whether or not Boneh-Durfee's attack suffers from the same problem before reaching $\delta=.292$.  See the end of Section~\ref{sec:overview} for more comments.

On the practical side, one sees another obstacle from (\ref{bdbd}):~attacks which require enormous lattices are infeasible  and cannot come close to their theoretical limitations anyhow.  Our approach here is to develop a method which gives experimentally verified improvements over previous work.
}  the bound (\ref{bdexp}), which still stands as the current record despite many attempts to improve it, or to even rigorously establish the needed algebraic independence (see \cite{bauerjouxarticle}).

Since the LLL algorithm has a widespread reputation for outperforming its provable guarantees, one might surmise that the bound (\ref{bdexp}) is more modest than actual experiments would indicate.  Surprisingly, the opposite is true:~all successful experiments in the literature work only for  $\delta$ relatively far below the theoretical upper bound of $1-2^{-1/2}\approx .292$ \cite{BD,BM,Wong}.  The reason for this is that (\ref{bdexp}) is an asymptotic estimate that requires very large lattices.
   Specifically, for 1,024-bit moduli $n$
  \begin{equation}\label{bdbd}
  \gathered
    \text{ Boneh-Durfee's attack requires lattices} \\ \text{of dimension $\ge 500$ for  $\delta\ge .278$.}
    \endgathered
  \end{equation}
   The difficulty of finding short vectors in lattices of dimension $\ge 500$ already serves as the hard underlying problem behind other cryptosystems, and indeed $\delta=.278$  is close to the limit of known experiments in \cite{BD,BM,Wong}.\footnote{It follows that it is important to distinguish between theoretical ranges of applicability of Coppersmith's method, and the practical ranges that they could possibly be applied to (in light of the difficulty of lattice basis reduction in large dimensions).}

Thus implementing lattice-based attacks can itself face impractically difficult problems even in ranges covered by theoretical guarantees.  It is therefore natural to ask the following questions:
\begin{enumerate}
  \item[{\bf Q1.}] Can one practically solve small-exponent RSA instances for $\delta$ significantly larger than the experiments reported in \cite{BD,BM,Wong}?
  \item[{\bf Q2.}] Is there a barrier from algebraic independence that creeps in before the theoretical upper bound is reached?  If so, how does one estimate the true range of validity of the attack?
  \item[{\bf Q3.}]
 How can Coppersmith's method be modified to reduce the size of the matrices (and their entries) involved?  It has long been considered to look at sublattices, but what are the optimal sublattices to choose?
\end{enumerate}

In  \secref{sec:overview} we review Coppersmith's method and the Boneh-Durfee attack, and  comment  on some nuances of comparing theoretical analyses to actual outcomes in applications of Coppersmith's method (e.g., {\bf Q2}).
The main contribution of this paper is to {\bf Q3}, by introducing a method in \secref{sec:paring} (influenced by the idea from machine learning of trying to find patterns in known examples, rather than being guided solely  by theory) to cut down the matrix size and hence   push back the choke point that high dimensional lattice basis reduction algorithms face in practice.  We use that to address {\bf Q1}   in \secref{sec:experiments1}, where we show our method is faster (and often more effective).

   In \secref{sec:experiments2} we present another example of how the ``focus group'' method can be applied to the attack in \cite[\S6]{wild}, where the size of the spanning set can be reduced significantly.  We chose these two examples because they are prominent theoretical and practical applications of Coppersmith's method to RSA.
 All the computations here (unless otherwise noted) were performed in Mathematica\footnote{Specifically, using Mathematica's {\tt{LatticeReduce}} command, which implements the L2 algorithm of Nguyen and Stehl\'e \cite{NS}.}  v.11  on a
  Dell PowerEdge R740xd server equipped with two Intel Xeon Silver 4114 2.2GHz processors and 256GB RAM.
In particular we did not use specialized lattice basis reduction packages such as \cite{fplll,ntl}.

We would like to thank Dan Boneh, Henry Cohn, Nadia Heninger, Jeff Hoffstein, Antoine Joux, Daniel Lichtblau, Alexander May, Oded Regev, Adi Shamir, Noah Stephens-Davidowitz, and David Wong for their helpful discussions.  We are also very appreciative of the anonymous referee for their very helpful comments, and to Galen Collier and the staff of the Rutgers Office of Advanced Research Computing for their assistance with Rutgers' Amarel high performance cluster.

\section{An overview of Coppersmith's method and Boneh-Durfee's attack on RSA}\label{sec:overview}

As before, let $p$ and $q$ be secret large prime numbers of comparable size, and $n=pq$ the public RSA modulus.  Let $e$ be the public encryption exponent and $d=n^\delta$ be the secret decryption exponent, which
satisfy  $ed\equiv 1\imod{\phi(n)}$, where $\phi(n)=(p-1)(q-1)=n-p-q+1$.
 In this case $d$'s relation to $e$ can be restated as the existence of an integer $k$ such that
\begin{equation}\label{rsak}
  e\,d \ \ = \ \ 1 \ + \ k\,\phi(n)\,, \ \ \text{where} \  k\,\approx  \,n^\delta\,,
\end{equation}
in which we have made the natural -- and trivially verifiable --  assumption that the public exponent $e$ has comparable size to $n$.  After dividing both sides by $d\phi(n)$ and using the fact that $n-\phi(n)=O(\sqrt{n})$, this implies
\begin{equation}\label{greatapprox}
\aligned
  \left|
  \f{e}{n} - \f{k}{d}
  \right| \ \  & \le \ \ \left|
  \f{e}{n} - \f{e}{\phi(n)}
  \right| \ + \ \left|
  \f{e}{\phi(n)} - \f{k}{d}
  \right| \\
 & = \ \ O\(\f{e}{n^{3/2}}\) \ + \ \f{1}{|d\phi(n)|} \ \ = \ \ O(n^{-1/2})\,.
\endaligned
\end{equation}
Wiener \cite{wiener} observed that if $\delta<\f 14$, the fraction $\f kd$ approximates $\f en$ much more accurately than $d^{-2}\gg n^{-2\delta}$, which is an unusually good approximation of a real number by a rational number of denominator $d$.  Hence $\f kd$ occurs among the continued fraction approximants to $\f en$, and  can be very efficiently computed.

Following \cite{BD}, consider the bivariate polynomial
\begin{equation}\label{fdef}
  f(x,y) \ \ = \ \ x(n-y) \ + \ 1  \,,
\end{equation}
which according  to (\ref{rsak}) satisfies
\begin{equation}\label{fmode}
  f(x_0,y_0) \ \ \equiv \ \ 0 \ \imod e\, ,
\end{equation}
where
\begin{equation}\label{x0y0}
  \ x_0 \ = \ k \ = \ O(e^\delta) \ \ \ \  \text{and} \ \ \  \
  y_0 \ = \ n-\phi(n) \ = \ O(\sqrt{e})
\end{equation}
are both relatively small compared to the modulus $e$.
Coppersmith's method (in this example, following Howgrave-Graham \cite{H-G}) is used to promote a polynomial congruence  relation such as  (\ref{fmode}) into a system of two    integer polynomial equalities, which can then be solved using classical methods.  To illustrate this in terms of the Boneh-Durfee attack,
define $x$-shifts and $y$-shifts
\begin{equation}\label{bdshiftpolys}
\aligned
  g_{i,\ell,m}(x,y) \  \ & = \ \ x^i\,f(x,y)^\ell\,e^{m-\ell} \\
  \text{and} \ \ \ h_{j,\ell,m}(x,y) \  \ & = \ \ y^j\,f(x,y)^\ell\,e^{m-\ell}\,,
\endaligned
\end{equation}
for
\begin{equation}\label{bdchoices}
 0\le \ell \le m\,, \ 0\le i \le m-\ell \, , \ \ \text{and} \ \ 1\le j \le t\,.
\end{equation}
They satisfy
\begin{equation}\label{modem}
  g_{i,\ell,m}(x_0,y_0) \ \ = \ \ h_{i,\ell,m}(x_0,y_0) \ \ \equiv \ \ 0 \imod{e^m}
\end{equation}
and span a sublattice $\L$ of $\R[x,y]$, the latter of which is endowed with the sum-of-squares norm $\|\cdot\|$ on polynomial coefficients. A short vector in this sublattice is a polynomial with small coefficients, ideally small enough that its value at a particular point such as $(x_0,y_0)$  will itself be relatively small.  By (\ref{modem}), that value is also a multiple of $e^m$; thus if it is less than $e^m$ in absolute value, it must actually vanish.

To make this more precise in our setting, let $X$ and $Y$ be upper bounds for $|x_0|$ and $|y_0|$, respectively (such as provided in (\ref{x0y0})).  Howgrave-Graham \cite{H-G} observed that if a polynomial $p(x,y)\in\L$ satisfies\footnote{Note that vector length $\|\cdot\|$ is not necessarily the appropriate metric.  The length condition (\ref{hg}) is quadratic in the polynomial coefficients, but the actual value of interest (the polynomial evaluated at a particular point) is instead linear:~it is the value of a linear functional on $\Lambda$.  Of course bounding the norm bounds the value of a linear functional, but possibly with a significant loss.  One might imagine leveraging some geometric information about $x_0$ and $y_0$ (such as their sign) which is known in advance.}
\begin{equation}\label{hg}
 \|p(xX,yY)\| \ \ < \ \ \frac{e^m}{\sqrt{w_p}}\,,
\end{equation}
  where $w_p$ is the number of nonzero monomials in $p(\cdot,\cdot)$, then an application of Cauchy-Schwartz shows   $|p(x_0,y_0)|<e^m$.  In particular, $(x_0,y_0)$ is a root of $p(\cdot,\cdot)$ over $\Z$ since   $p(x_0,y_0) \equiv 0\imod{e^m}$. Boneh-Durfee prove that this norm condition is met for the shortest   vector outputted by LLL provided
\begin{equation}\label{BDenabling}
  |\L| \ \ \le \ \ e^{m(w-1)}(w2^w)^{(1-w)/2}  \, , \ \ w \,=\,\dim(\L)\,,
\end{equation}
where $|\L|$ denotes the covolume of $\Lambda$.
For $\delta<\f{7}{6}-\f{\sqrt{7}}{3}\approx.284$ this ``enabling'' condition is met for sufficiently large values  of $m$ and $e$.  We shall refer to this as the Boneh-Durfee ``.284'' attack, in order to distinguish it from their more refined analysis (using a carefully selected sublattice) that extends the range to $\delta<1-2^{-1/2}\approx .292$.  See also \cite{BM,HM,KSI} for other attacks theoretically deriving exponents of this size, or close to it.

Under the enabling condition (\ref{BDenabling}), the two shortest vectors outputted by LLL are  coefficients of bivariate polynomials which vanish at $(x_0,y_0)$.
Boneh-Durfee point out that in practice these polynomials are algebraically independent, and thus their common roots can be extracted using resultants.  However, this observed algebraic independence has not yet been rigorously established (see \cite{bauerjouxarticle}).  In principle (as can happen, for example, if one does not use any of the polynomials $h_{j,\ell,m}$) the shortest vectors may all result in polynomials which are trivial multiples of each other, and hence not give enough equations to reveal the two unknowns $x_0$ and $y_0$.
\\

\noindent{\bf An interesting example:}
 consider the 1,000 bit RSA modulus $n=pq$ and key $d=n^\d$ given by
\\

$p=327534248375076317083641611376534056264358811260976111
454743\\~~~~~~~~~~~~~~~\,469579874653650577266211366585026890270802159105074832098421\\~~~~~~~~~~~\,5116927258714434174724054953133\,,$\\

$q=327462704072360233831723075103626846066746692190298143145154\\~~~~~~~~~~~~~~~\,087005180715732984190358817594057449905
589163120424047417288\\~~~~~~~~~~\,34002393744713793935
71624577657\,,$\\
%
%$e=361363878268493379275397541299165218700869230719585639806\\~~~~~~\,9061786869922
%87593014458126635041612888998868994168084520\\~~~~~~\,  16722338275914262235205169
%4686280109921416903394682744927\\~~~~~~\,   512263566493840489580120544105203367036
%226838000325631803\\~~~~~~\, 4307687881649513249906298975908193029604188253736602
%14656\\~~~~~~\,5547861332549763\,,$

\noindent and\\

$d=300147077152565471186517713474704374146330287118250537992743\\ ~~~~~~~~~~~ \,5326735028
048350149451\,, \ \ (\delta \approx .2707).$\\

\noindent  We applied the  BKZ lattice basis reduction from \cite{ntl} with block size 3 to the lattice from
Boneh-Durfee's .292 attack with parameters $(m,t)=(5,2)$.
Of the 25 output vectors, only one of them (the fifth longest\footnote{A similar  feature was already observed in A.~Bauer's Ph.D. thesis \cite{bauerthesis}.}!)     produces a polynomial which vanishes at $(x_0,y_0)$.  Interestingly and perhaps counterintuitively, applying BKZ with larger block size (such as 5) failed to produce {\em any} vectors vanishing at $(x_0,y_0)$.  (We were unable to do any better using the implementation of BKZ in {\tt sagemath}.)  This example demonstrates that the vector length of the output basis is not the sole determinant of success, in addition to the  sensitivity to the choice of lattice basis reduction method.
\\

It is worth mentioning other lattice attacks which use     polynomials different from (\ref{fdef}), and which also consistently beat Wiener's $\delta<1/4$ bound. Bauer's thesis
\cite[Chapter 4]{bauerthesis} discusses a three-variable analog based on using a short continued fraction approximation of $e/n$, stopping roughly at the point at which it is theoretically expected to differ from that of $e/\phi(n)$.  Two additional integer parameters are then substituted to account for the remaining part of the continued fraction approximation.\footnote{See also the paper \cite{Dujella}.  The anonymous referee of the present paper kindly suggested another approach:~attempt to guess a portion of the bits of these unknown parameters, in the hopes of gaining a more-than-compensating savings from the smaller lattices which result from analyzing the remaining portion.}   A lattice is again formed as above using  congruences modulo powers of $e$.  Her analysis of the enabling condition shows that when $\delta < .34$, the lattice has short vectors that produce polynomials which vanish at the desired roots.   Alternatively, one can instead apply the lattice method of \cite{JM} to this partial continued fraction approach (as we have attempted in experiments) -- its analogous enabling condition holds for $\delta<1/3$.  These are theoretical ranges in which Coppersmith's method will provably find short vectors (for large enough lattice sizes), yet possibly nevertheless fail to factor $n$ because of algebraic dependence.  Both ranges extend much further than Boneh-Durfee's $\delta < 1-2^{-1/2}\approx .292$ range, and the lattice sizes in both attacks can be   improved using the ``focus group'' methodology in \secref{sec:paring} below.

Indeed, despite the promising increase in this range for  $\delta$ from .292 to $1/3$ or $.34$, neither of these approaches comes near .292 in practice.
The results of our limited experimental trials  indicate that the actual performance of either of these algorithms seems roughly  comparable to that of Boneh-Durfee's.
  In particular, the experiments show that algebraic independence fails at a much earlier point, well before the enabling condition of $\delta<1/3$ or $\delta<.34$ is reached.   Furthermore,  the lattice sizes necessary to study such large exponents $\delta$ are themselves impractically large.  That calls into question the direct practical relevance of the enabling condition itself, and demonstrates the importance of a better understanding of the actual performance of these attacks.\\

\noindent {\bf Remarks on sublattices and minimizing $|\L|$:}~in order to leverage provable guarantees that a lattice basis reduction algorithm will find a sufficiently short vector, sublattices in variants of Coppersmith's attack are often taken in order   to  effectively reduce the  covolume  $|\L|$ (see {\bf Q3} in Section~\ref{sec:intro}).   This is primarily done to ensure the validity of an enabling condition such as (\ref{BDenabling}).   While this allows for rigorous, theoretical analysis, there are geometric reasons why  it may not be optimal:
\begin{itemize}
\item If $\Lambda$ does not behave like a random lattice, it may have vectors at several length scales that do not interact much with each other.
\item For example, suppose one appends  a very long vector to a short basis of a lattice perpendicular to it.  This would magnify the covolume without affecting the outcome of lattice basis reduction at all.
\item The ultimate goal in Coppersmith's method is not to  reduce the covolume, but to increase the likelihood of finding a short vector.  It is more important to   identify sublattices having short vectors, which is not well-measured by the covolume.
\item As we have noted in the above discussion of  Bauer's thesis \cite{bauerthesis}, attacks with very different enabling condition bounds may perform similarly in practice, since algebraic dependence may creep in before the enabling condition is reached.  Thus $|\L|$ itself may not actually enter into a meaningful bound anyhow.
\end{itemize}

\begin{center}
\begin{figure}
\includegraphics[width=2.24in]{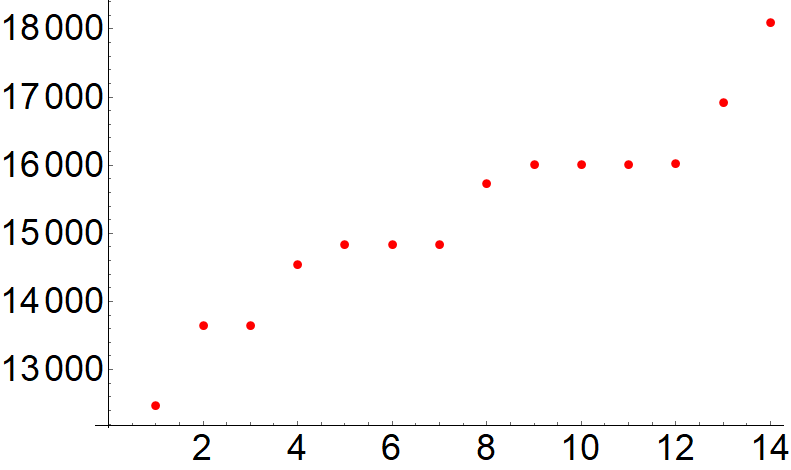}
\hspace{.2in}
\includegraphics[width=2.24in]{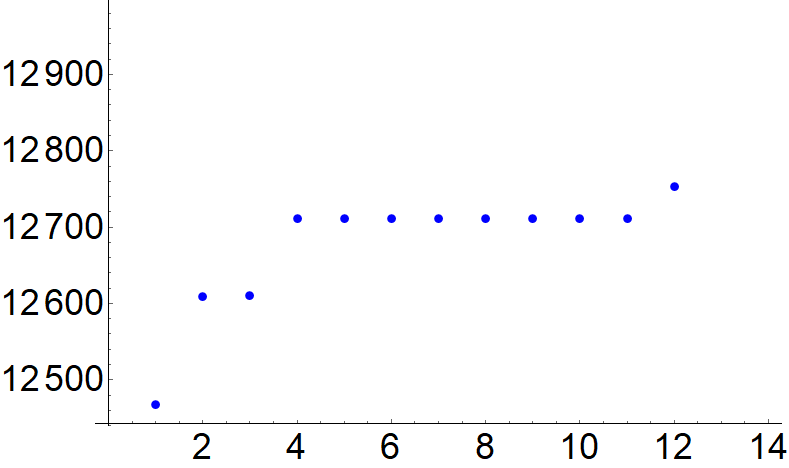}
\caption{Natural logarithms of lengths of lattice basis vectors in the Boneh-Durfee .284 attack with $n\approx 2^{6,000}$ and $\delta\approx .284$, before and after lattice basis reduction.  The attack failed in this instance.  \label{fig:clumps}}
\end{figure}
\end{center}

\noindent
We conclude this section by remarking  that the lattices produced in the Boneh-Durfee attack appear to be far from random, as is evidenced by their vector lengths.  This appears to be in contrast with the lattices produced in other applications of Coppersmith's method -- though not all (e.g., \cite{CH}).  For example, one typically expects a basis outputted by the LLL algorithm \cite{LLL} to have vectors of comparable length.
 Figure~\ref{fig:clumps} shows the logarithms of the vector lengths in the original and reduced lattice bases for an instance of the Boneh-Durfee .284 attack with $n\approx 2^{6,000}$ and $\delta\approx .284$.
At this logarithmic scale one can see clumps of basis vectors of roughly the same length, yet nevertheless the overall lengths of the basis vectors do differ significantly within each plot.  The plot the left indicates that the input basis has several different regimes, owing to the structure of (\ref{bdshiftpolys})-(\ref{bdchoices}).  The plot on the right shows that the output basis also has vectors in (fewer) clumps of similar logarithmic length, in particular with a large separation between the shortest vector (which represents a constant polynomial)  and the others.   Not surprisingly, the attack failed in this particular instance.  Understanding this ``clumping'' phenomenon may help gain insight into Coppersmith's method.    For example, is there inhomogeneity in the geometry of the lattice that effectively reduces its dimension?  If so, can it be exploited? This is the underlying geometric motivation behind the ``focus group'' method in \secref{sec:paring}, which is an approach to identifying sublattices having short vectors.

\section{``Focus group'' attacks}\label{sec:paring}

Applying lattice basis reduction to a sublattice may increase the chances of finding short vectors, while of course simultaneously decreasing  run time.  Given the inherent limitations in performing lattice basis reduction in high dimensions, sometimes finding an appropriate sublattice can make the difference between finding solutions and finding no solutions at all (or not even being able to fully execute lattice reduction).

In this section we describe a principled, evidence-based approach to selecting a sublattice in certain lattice basis reduction problems, such as applications of Coppersmith's method.  Its main idea is to deform to a simpler problem in which one can directly determine which basis vectors contribute nontrivially to the shortest vectors.  This methodology is applied in  \secref{sec:experiments1} to small-exponent RSA and in \secref{sec:experiments2} to the ``Coppersmith in the wild'' smart card attack of \cite{wild}.

This ``focus group'' attack consists of three main steps:
\begin{enumerate}
\item  {\bf Set small parameters.}  Find a regime with the same lattice dimension, but reduced sizes of basis vectors coordinate entries.  This  makes it faster (or even possible) to execute lattice basis reduction on large matrices.  For example, in the case of small-exponent RSA we set $\delta$ in (\ref{smallexp}) to be slightly larger than $\f 14$ (which is the point at which Wiener's continued fraction attack ceases to work).  Of course would be desirable to improve this step by theoretically understanding in advance which basis vectors to keep (e.g., using the notion of ``helpful vector'' from \cite[Chapter 7]{Maythesis} rather to rely on experiments), but this may   not be practical because of the complexity of lattice basis reduction.

\begin{figure}
\begin{center}
\includegraphics[width=2.8in]{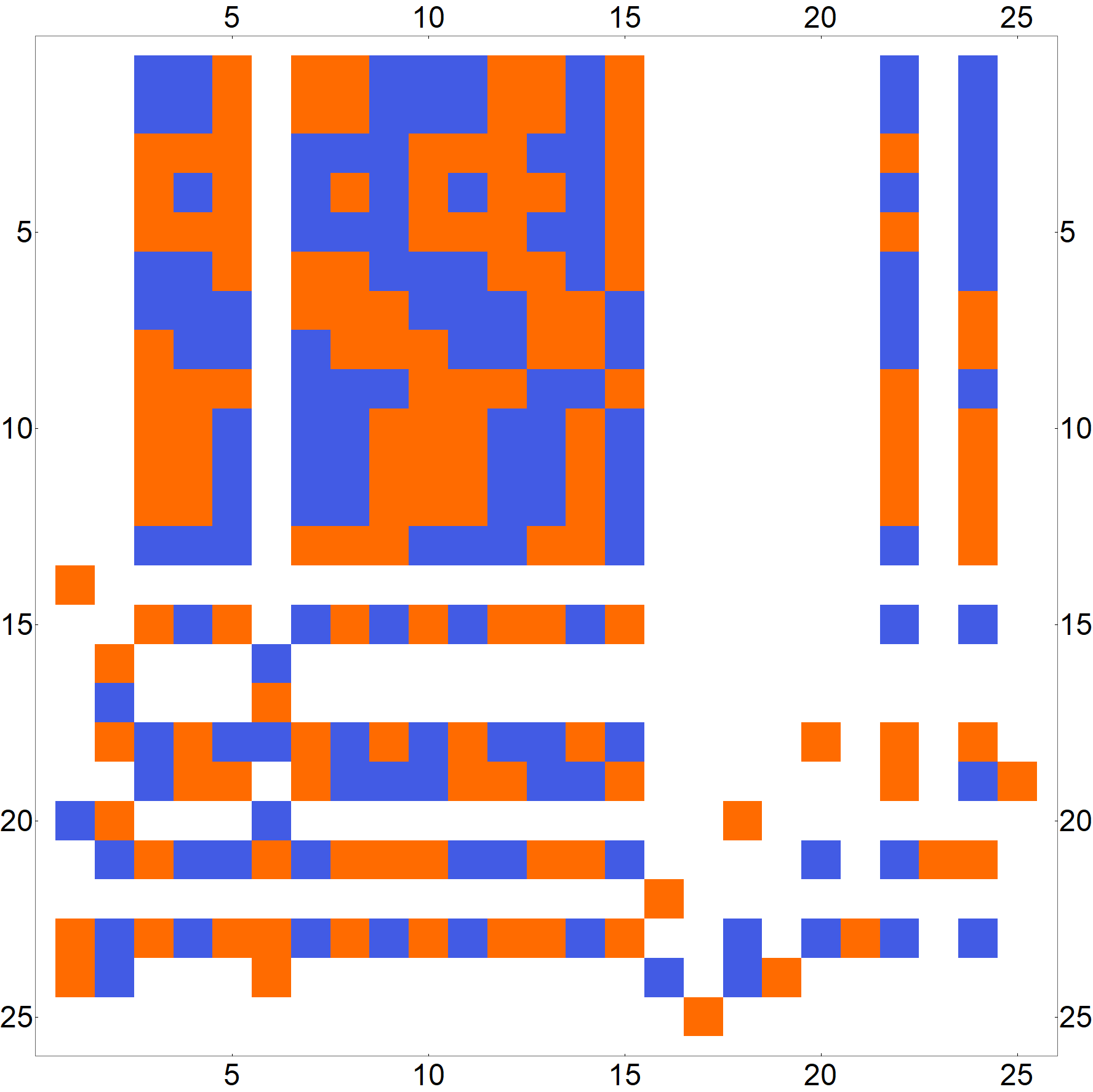}
\caption{  A representation of the change of basis matrix for  the    lattice basis reduction step in Boneh-Durfee's .284 attack (see the text for more details).  The matrix has a number of columns with many zero entries (marked white).\label{fig:changeofbasis}}
\end{center}
\end{figure}

\item {\bf Check the output to see which parts of the original basis were actually used.}
     Figure~\ref{fig:changeofbasis}
pictorially represents the change of basis matrix for the    lattice basis reduction step in Boneh-Durfee's .284 attack for a 6,000-bit RSA modulus $n$, with $\delta\approx .251$  and parameters $(m,t)=(4,2)$ (see (\ref{bdchoices})).  The columns are indexed by the input basis vectors and the rows are indexed by the output basis vectors.  Each entry in the matrix is plotted as  blue/dark gray (positive), orange/light gray (negative), or white (zero).

 The long white vertical streaks emanating from the top of the figure reveal that certain input basis vectors  are not used in forming the shortest vectors in the lattice output.  Those basis elements from (\ref{bdshiftpolys})-(\ref{bdchoices}) can be graphically represented as in Figure~\ref{fig:dots},
where the figure on the left represents the $x$-shifts and the figure on the right represents the $y$-shifts.  Here the unfilled white circles indicate unused vectors and filled black circles indicate useful vectors.  Boneh-Durfee's .292 attack refines their .284 attack by discarding  some $y$-shifts from (\ref{bdchoices}), but not the same ones as here (theirs are chosen to minimize the covolume $|\L|$).  In fact, the figure indicates that most of the $y$-shifts are not actually used.  It is more striking that some of the smaller $x$-shifts are not used, consistent with the utility of a similar device in \cite[\S4]{BM}.  Similar patterns arise for larger parameter sizes and were used to formulate the attack in Section~\ref{sec:experiments1}.  Indeed, examples of patterns yield useful descriptions in terms of extra additional parameters, which are then used to extrapolate good guesses for which families of sublattices to look at in more challenging situations.

\item {\bf Remove unused basis elements.} This has advantages  for run time, storage, and quality of results, since lattice basis reduction on smaller lattices typically performs dramatically better.  After all, the approximation factor in lattice basis reduction is tighter in smaller dimensions.
\begin{figure}
    \begin{center}
\includegraphics[width=4.7in]{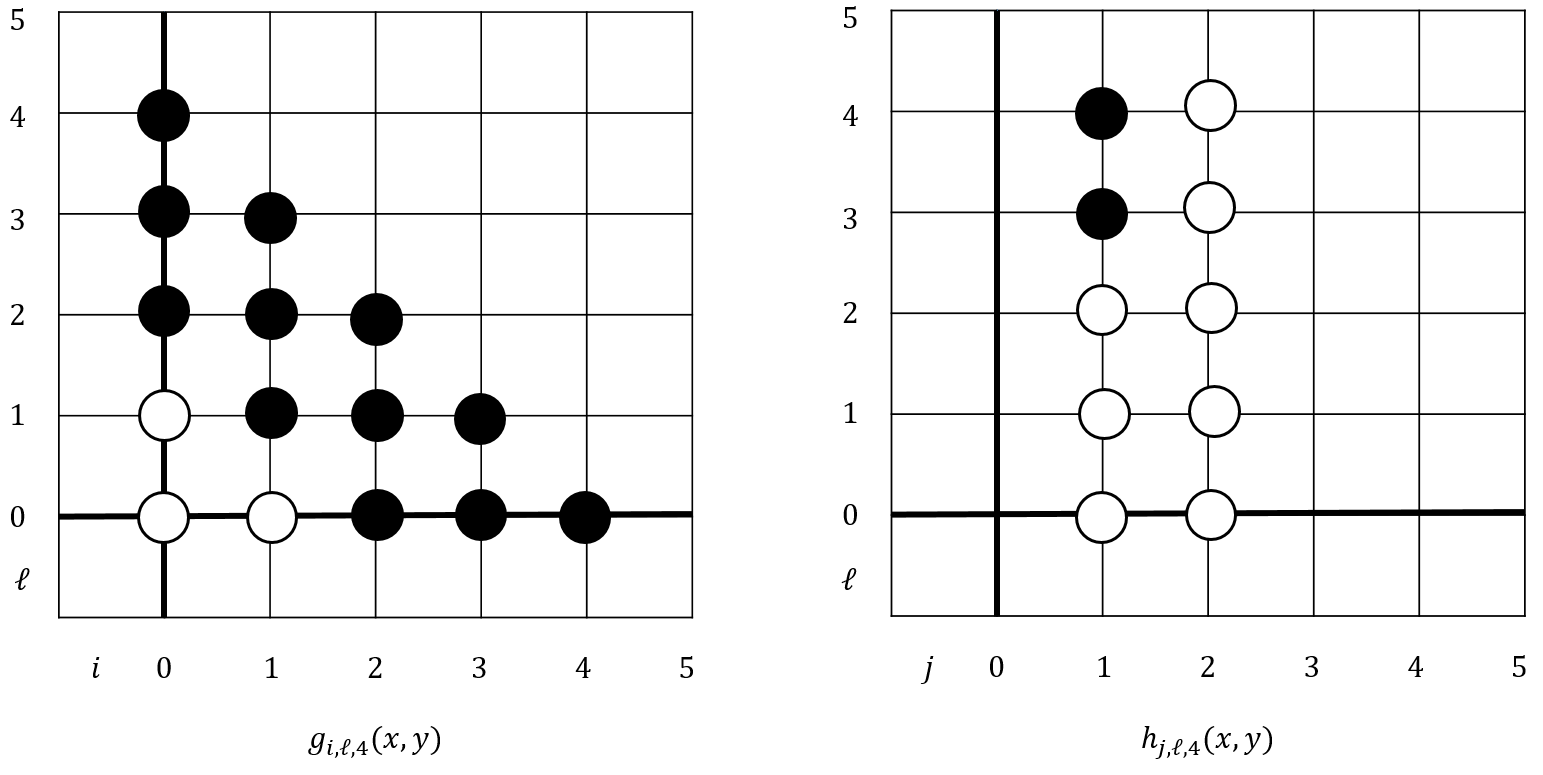}
\caption{A representation of which polynomials  in (\ref{bdshiftpolys})-(\ref{bdchoices}) are actually used (black circles) in forming the shortest vector in the lattice basis reduction step for a particular instance of the Boneh-Durfee attack.   The unfilled, white circles represent discarded basis vectors  (see the text for more details).
 \label{fig:dots}}
\end{center}
\end{figure}

\end{enumerate}

\section{The ``focus group'' attack on small-exponent RSA}\label{sec:experiments1}

We now specialize the methodology of \secref{sec:paring} to small-exponent RSA.  Trials of the Boneh-Durfee .284 attack \cite{BD} with small parameters suggest   a particular sublattice to use, which we shall describe below.      Previous work has selected sublattices using other methods.  For example,    Boneh-Durfee  suggest in their .292 attack to  remove certain $h_{j,\ell,m}$ which contribute large factors to the covolume.  Later work by Bl\"omer and May \cite{BM} suggests removing some of the $g_{i,\ell,m}$ as well (see also \cite{HM,KSI}).

  Our approach is   guided by which vectors are likely to contribute to a nontrivial solution, but not directly by determinant considerations.  We introduce two integer parameters  $\sigma$ and $\tau$ (in addition to $m$ and $t$),  and exclude from (\ref{bdchoices}) all indices with $i+\ell\le \sigma$ and $\ell-2j\le \tau$ (this is motivated by the shape of the black and white circles in Figure~\ref{fig:dots}).  That is,  the polynomials in (\ref{bdshiftpolys}) are taken for indices
\begin{equation}\label{newchoices}
0\le \ell \le m,\, \max(-1,\sigma-\ell)< i \le m-\ell,\, \text{~and~} 1\le j \le \min(t,1+\smallf{\ell-\tau}{2})
\end{equation}
instead of (\ref{bdchoices}).  We choose $X=\lceil 2e^\delta\rceil$ and $Y=\lceil 2e^{1/2}\rceil$ as rough integral upper bounds for $x_0$ and $y_0$, respectively (cf.~(\ref{x0y0})).

\subsection*{Experiments}

We ran timings using Mathematica v.11 on a Dell PowerEdge R740xd server with two Intel Xeon Silver 4114 2.2GHz processors and 256GB RAM.  We did not seriously attempt to optimize the lattice basis reduction computations, relying instead on Mathematica's {\tt LatticeReduce} command (which is an implementation of \cite{NS}).   Timing results are presented in Table~\ref{tab:tab1} and include, as a control experiment, a comparison with an implementation of Boneh-Durfee's  .292 and .284  attacks using Mathematica on the same machine with the same parameters $m$ and $t$.   It would be interesting to perform a similar comparison with the algorithm of \cite{BM}, whose sublattice is more similar to the one selected by the ``focus group'' attack.  The attack in \cite{BM} also satisfies the enabling condition for the same $\delta<1-\sqrt{1/2}\approx .292$ range as Boneh-Durfee's   attack \cite{BD}.  We have not rigorously analyzed at what point our enabling condition breaks down, as it may be moot anyhow:~algebraic independence might be lost before that point  (see the comments at the end of \secref{sec:overview}).

\begin{table}[p]
\scriptsize
\begin{tabular}{|c|r|c|c||r|c||r|c||r|c|c|}
  \hline
  % after \\: \hline or \cline{col1-col2} \cline{col3-col4} ...
   \multirow{2}{*}{bits of $n$}&  \multirow{2}{*}{trials} &
   \multirow{2}{*}{$\delta$}
  &
   \multirow{2}{*}{$(m,t,\sigma,\tau)$}
  & \multicolumn{2}{c|}{ Focus group} & \multicolumn{2}{c|}{Boneh-Durfee .292} & \multicolumn{2}{c|}{Boneh-Durfee .284} \\
     &     &     &  &  time & dim & time & dim & time & dim \\
  \hline
  1000 & 100 & .270 & (6,2,2,0)  & 8.46 s &28    & 15.95 s &34  &  30.29 s &42 \\
  \hline
  4000 & 100 & .273  & (6,2,2,0) & 60.37 s &28   & 103.49 s &34  &  182.92 s &42 \\
 \hline
  6000 & 15 & .277  & (8,3,2,-1) & 1905.81 s &54   & 2170.15 s &57 & 3920.45 s &72 \\
  \hline
  10000 & 100 & .260 & (3,1,1,0) & 1.70 s &14  & 2.76 s &17  & 4.47 s &20  \\
  10000 & 100 & .265 & (4,1,1,0)  & 12.94 s &14   & 17.90 s &17  & 24.76 s &20  \\
  10000 & 15 & .277 & (8,3,2,-1) & 4565.47 s  &54  & 5063.59 s &57   & 8492.21 s &72  \\
  \hline
\end{tabular}
\caption{Timings of trials of the ``focus group'' attack on small-exponent RSA.  Times listed are averages over many trials of RSA keys, in which each of the three attacks on the right is performed on the same key. Comparisons  are given in the last two columns for the Boneh-Durfee .292 and .284 attacks with the same values of $m$ and $t$.  We list the average time in seconds as well as the dimension of the lattices involved, which get progressively larger as one goes from the focus group attack to the Boneh-Durfee .284 attack.
 The focus group attack is significantly faster, and uses less memory due to the smaller lattice size.   (All computations were run on the same machine using the same lattice basis reduction algorithm.)
 Times refer to the lattice basis reduction step only.       \label{tab:tab1}}
\end{table}

\begin{table}[p]
\scriptsize
\begin{tabular}{|c|r|c|c||r|c||r|c||r|c|c|}
  \hline
  % after \\: \hline or \cline{col1-col2} \cline{col3-col4} ...
   \multirow{2}{*}{bits of $n$}&  \multirow{2}{*}{trials} &
   \multirow{2}{*}{$\delta$}
  &
   \multirow{2}{*}{$(m,t,\sigma,\tau)$}
  & \multicolumn{2}{c|}{Focus group} & \multicolumn{2}{c|}{Boneh-Durfee .292} & \multicolumn{2}{c|}{Boneh-Durfee .284} \\
     &     &     &  &   Success \% & dim &Success \%& dim & Success \% & dim \\
%
%
% & $\srel{\text{Focus group}}{\text{Success rate/dim}}$& $\srel{\text{BD .292}}{\text{Success rate/dim}}$ & $\srel{\text{BD .284}}{\text{Success rate/dim}}$ \\
  \hline
  1000 & 100 & .270 & (6,2,2,0)  & 100\% &28    & 100\% &34  &  100\% &42  \\
  1000 & 100 & .273 & (6,2,2,0)  & 43\% &28   & 64\% &34   &  64\% &42  \\
    1000 & 100 & .277 & (8,3,2,-1) & 21\% &54   & 10\% &57  & 9\% &72   \\
  1000 & 100 & .279 & (10,4,2,-3) & 62\% &92  & 26\% &85  & 26\% &110  \\
  1000 & 100 & .280 & (10,4,2,-3) & 1\% &92   & 0\% &85   & 0\% &110  \\ \hline
    4000 & 100 & .273 & (6,2,2,0)  & 58\% &28   & 100\% &34  &  100\% &42  \\
    \hline
      6000 & 15 & .277  & (8,3,2,-1) & 100\% &54   & 100\% &57  & 100\% &72  \\
      \hline
  10000 & 100 & .260 & (3,1,1,0) & 100\% &14    & 100\% &17    & 100\% &20    \\
  10000 & 100 & .265 & (4,1,1,0)  & 100\% &14   & 100\% &17    & 100\% &20    \\
  10000 & 15 & .277 & (8,3,2,-1) & 100\% &54    & 100\% &57    & 100\% &72    \\
  \hline
\end{tabular}
 \caption{Success rates of the three algorithms in Table~\ref{tab:tab1}, as measured by producing multiple polynomials which vanish on the secret key (as a time-saving proxy to allow for more experiments to be run).  Again, each trial involves the three attacks on the right applied to the same RSA key.  The focus group attack has a lower success rate than Boneh-Durfee's attacks for some smaller values of $\delta$, but appears to outperform the Boneh-Durfee attack for   larger values of $\delta$ and lattice dimensions. More data is shown in the plots in  Figure~\ref{fig:success4000bit}. \label{tab:tab2}}
\end{table}

\begin{figure}[p!]
    \begin{center}
\includegraphics[width=5in]{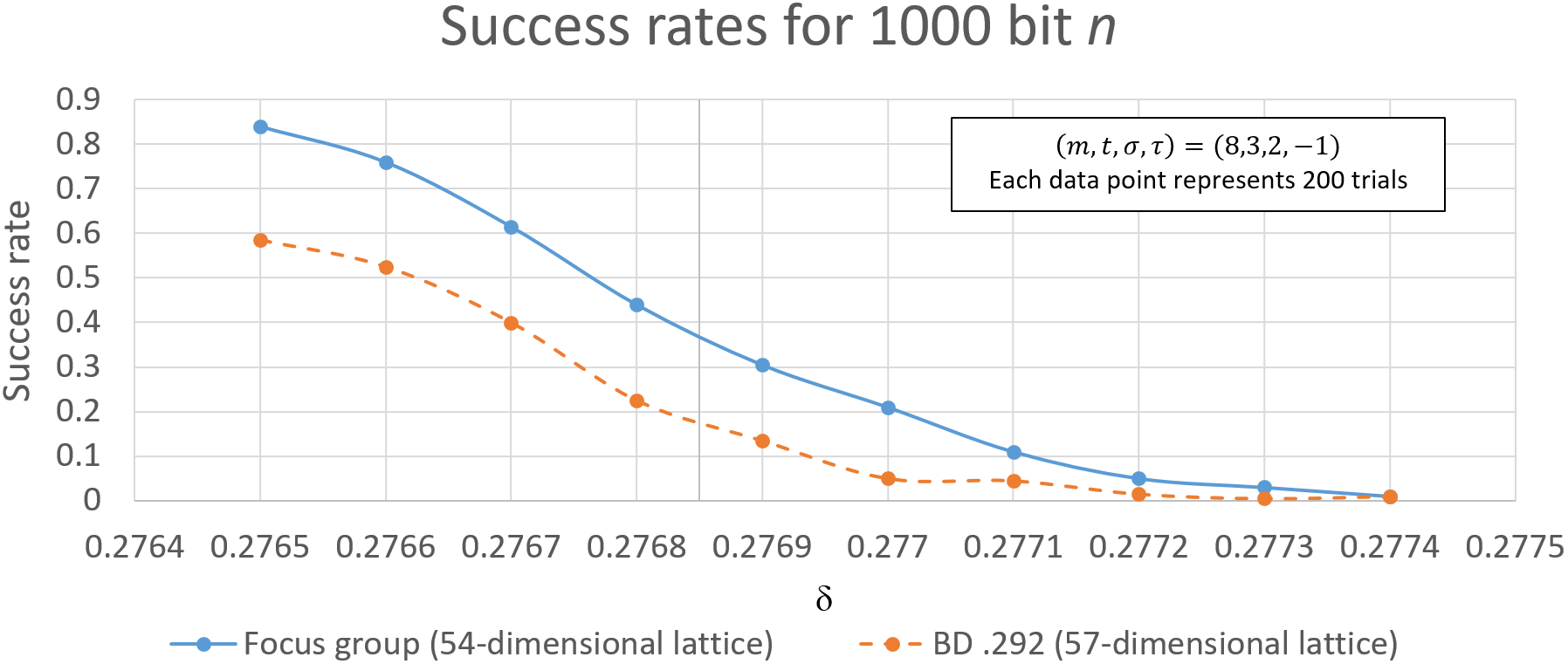}
\\ \vskip .3cm
\includegraphics[width=5in]{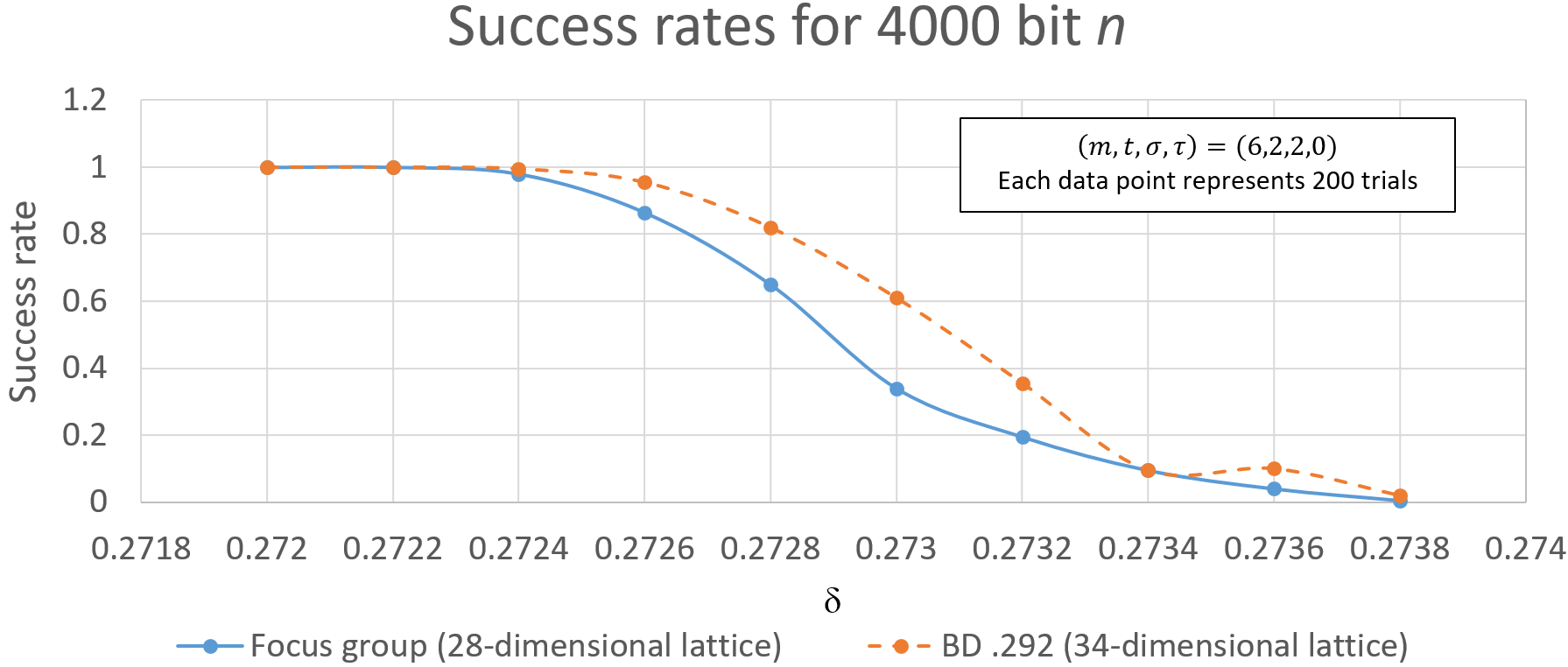}
\\ \vskip .3cm
\includegraphics[width=5in]{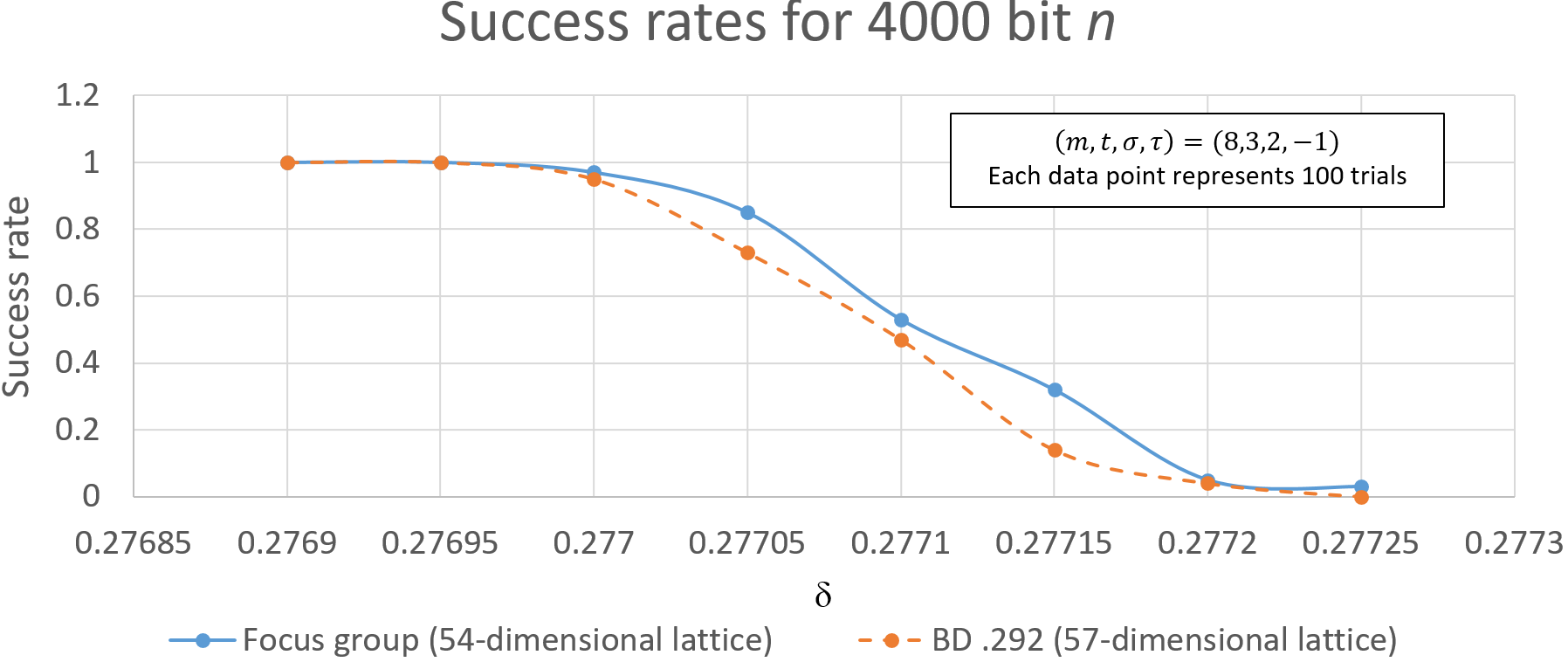}
\caption{These plots amplify the data in Table~\ref{tab:tab2} and show how the focus group method compares to the Boneh-Durfee .292 attack near their  limits of failure for larger exponents $\delta$.   It appears the focus group method outperforms the Boneh-Durfee .292 attack for larger  $\delta$ and lattice dimensions.
 \label{fig:success4000bit}}
\end{center}
\end{figure}

Extraction of the secret key after lattice basis reduction can be fairly slow, but we always found it to  be possible so long as more than one output polynomial vanishes on the secret key.  In order to generate more data on the success of the method and avoid this bottleneck, we ran a number of experiments in which success was measured by finding more than one such polynomial.  (In order to be more confident that no algebraic dependence issues arose, we performed a more rigorous analysis on several sample lattice reduction outputs and found sufficient algebraic independence in all cases.)
 Table~\ref{tab:tab2} and Figure~\ref{fig:success4000bit} show the probability of success for the focus group method compared to the Boneh-Durfee attacks.  These experiments were run on many machines in Rutgers' Amarel high performance cluster. For some parameter choices $(m,t,\sigma,\tau)$, the focus group attack is more successful than the Boneh-Durfee attacks (e.g., in the first plot in Figure~\ref{fig:success4000bit}), but not for all.  For example, in the entry for 4000 bit moduli $n$ with $(m,t,\sigma,\tau)=(6,2,2,0)$ and $\delta=.273$, the Boneh-Durfee attack was successful  on 100 out of 100 randomly chosen RSA keys, whereas the focus group attack was successful only 58 of those same 100 RSA keys.  The second plot in Figure~\ref{fig:success4000bit} shows this holds true for the same choice  of $(m,t,\sigma,\tau)=(6,2,2,0)$ and for exponents $\delta$ in a nearby range.  However, it appears this phenomenon may be limited to smaller $\delta$ and matrix sizes:~when one instead considers the larger lattices with $(m,t,\sigma,\tau)=(8,3,2,-1)$ (the last plot in  Figure~\ref{fig:success4000bit}), the situation apparently reverses and the focus group method is more successful than Boneh-Durfee's attack.  Table~\ref{tab:tab2}'s entries for 1000 bit $n$ and the first plot in Figure~\ref{fig:success4000bit} also suggest that the focus group may be more successful than Boneh-Durfee's attack in runs that involve larger lattices and exponents $\delta$.  We should caution, however, that we have not systematically analyzed this beyond the experiments presented here.
We are also unsure exactly what to attribute this apparent improvement to, though we suspect it is that lattice basis reduction algorithms have superior performance on smaller lattices.

As an aside, the timings demonstrate the power of the implementation of the L2 lattice basis reduction algorithm \cite{NS} used in Mathematica's {\tt LatticeReduce} command, which typically outperformed the BKZ implementations in NTL and {\tt sagemath}.  For example, the exponents $\delta$ achieved here are higher than in previously reported experiments (e.g., \cite{BD,BM,Wong}), with
the size of $d$ in the last entry in Table~\ref{tab:tab1} being 220 bits longer than achieved in \cite{BD} for a 10,000-bit RSA modulus $n$.  Recall that Table~\ref{tab:tab1} makes a controlled comparison between different approaches to Coppersmith's method on the same RSA keys, by using the same hardware and same lattice basis reduction algorithms.
 It is interesting to speculate whether certain features of \cite{NS} are particularly useful when applied to the lattices produced by Coppersmith's method, and if so, how to leverage them further (recall also the example in \secref{sec:overview}, where BKZ is more helpful with block size 3 than block size 5).

\section{The ``focus group'' attack applied to partial-key-recovery methods of \cite{wild}}\label{sec:experiments2}

In this section we apply the ``focus group'' methodology to the ``Coppersmith in the wild'' partial-key-recovery attack of \cite[\S6]{wild}, which   as far as we are aware was the first Coppersmith-style attack  successfully applied to real-world keys.  In their situation, $n=pq\approx 2^{1024}$ is an RSA modulus, and $p\approx 2^{512}$ has the special form
\begin{equation}\label{wild1}
  p \ \ = \ \ a \ + \ 2^t s \ + \ r\,,
\end{equation}
where $a$ and $t$ are -- or at least suspected to be -- known, but $s$ and $t$ are unknown.   The authors consider the polynomial $f(x,y)=a+2^t x+y$, and form  the ${k+2\choose 2}$-dimensional lattice spanned by
the polynomials
\begin{equation}\label{wild2}
  \{x^i y^h f(x,y)| 0\le i+h\le k-1\} \ \cup \  \{x^j y^h n| 0\le j+h\le k\}
\end{equation}
for an integer parameter $k\ge 0$.  Lattice basis reduction is then used (if successful) to obtain  two small algebraically independent polynomials, from which the values of $s$ and $t$ are extracted.

\begin{figure}[h]
    \begin{center}
\includegraphics[width=4.7in]{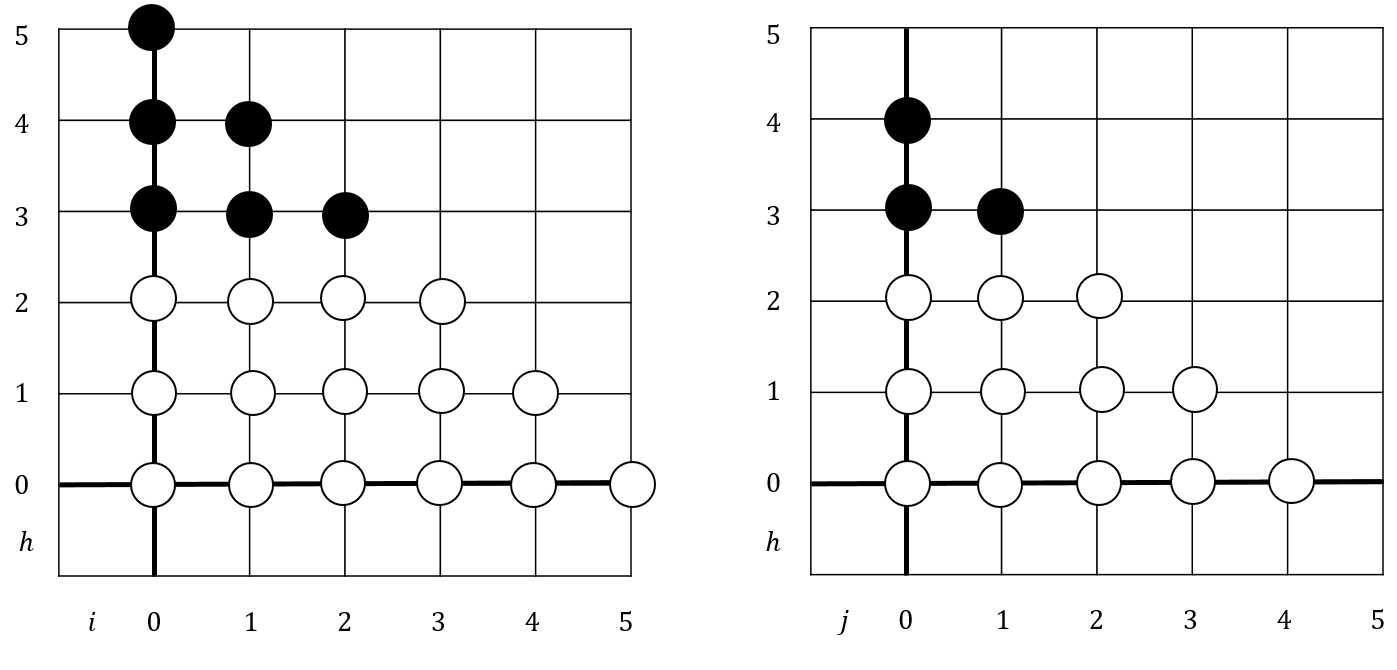}
\caption{A representation of which polynomials from (\ref{wild2}) are actually used (black circles) in solving for $s$ and $t$ in small examples. In this example $k=5$ and $h$ in (\ref{wild2}) is at least $h_0=3$.
 \label{fig:dotswild}}
\end{center}
\end{figure}

We performed experiments with one particular set of parameters successfully studied in \cite[\S 6.2]{wild}, namely $k=4$, $t=428$, $a=2^{511}+2^{510}$, $X=2^{100}$, and $Y=2^{28}$.  Experiments with small parameters revealed that solutions are often found within sublattices generated by the polynomials in (\ref{wild2}) having $h\ge h_0$, where $h_0$ is fixed (see Figure~\ref{fig:dotswild}).\footnote{After removing a common factor of $y^{h_0}$, the generators of this sublattice are given by (\ref{wild2}) but with $k$ replaced by $k-h_0$.  This might explain the remark on \cite[p.~355]{wild} that some experiments for small $k$ worked in situations where theoretically no useful output was expected.}  Following the focus group methodology,
we considered such a sublattice in this particular example with $h_0=1$, and verified that by taking these ten polynomials (instead of the 25 in (\ref{wild2})), we could just as easily recover the factorization of $n$ using lattice basis reduction applied to the corresponding sublattice.  Thus the same performance in this example is obtained from using this chosen ten-dimensional sublattice of the ambient 25-dimensional lattice.

{\bf The use of Hermite normal form in the ``focus group'' method:}
Hermite normal form has previously been applied to Coppersmith's method (e.g., \cite[Example 19.3.2]{Ga}).  For example, one can obtain the Hermite normal form of matrix whose rows form a generating set of the lattice (perhaps strictly larger than a basis), and then  apply lattice basis reduction  to only some of the rows of the  output.  Thus instead of using the generating set directly produced by   Coppersmith's method, we instead transform  it to a different generating set (using Hermite normal form) and apply the focus group methodology of selecting a sublattice at this secondary stage instead.

 This approach via Hermite normal form did not work well when applied to the previous example, but did find success when applied to primes (\ref{wild1}) of a different type than those studied in \cite{wild}.  Namely, here we
 arbitrarily chose  $a$ to be the smallest positive integer congruent to $3^{1,000}\pmod {2^{512}}$, and randomly generated $p$ of the form (\ref{wild1})  with $X=Y=2^{25}$, and $t=488$.  We took $k=5$ and studied the $36\times 21$ matrix of coefficients of the generating set (\ref{wild2}) with respect to the monomial basis $\{x^i y^h|0\le i+h\le k\}$, and considered the lattice spanned by the top ${k+1\choose 2}=15$ rows of its Hermite normal form as generated by Mathematica's {\tt HermiteDecomposition} command.  Lattice basis reduction applied to this 15-dimensional lattice then yielded polynomials which easily recovered the factorization of $n$.

\section{Conclusions} \label{conclusions}

We have considered the small-exponent RSA problem and attacks on it using Coppersmith's method, which relies on finding short vectors in a lattice.  Using theoretical and experimental observations, we have proposed a principled technique to restrict lattice basis reduction  to a carefully-selected sublattice, based on the behaviour of simpler examples.  This    ``focus group'' attack  specifically takes into account    which parts of the lattice are likely to be   used.  When applied to the  small-exponent RSA problem, it points to  a geometric structure of the lattice in Boneh-Durfee's attack \cite{BD} that can be leveraged to reduce the running time and memory usage of the lattice basis reduction step, while allowing the attack to be applied to more RSA keys.

Several interesting questions remain  for future investigations.  For example, Mathematica's implementation of the L2 \cite{NS} lattice basis reduction algorithm  accounted for a several hundred bit improvement in some experiments over previous work (which instead used LLL \cite{LLL}), an improvement which cannot be explained by hardware advances alone.  Is it possible that special features of the lattices generated by Coppersmith's method can be exploited by new, specially designed lattice basis reduction algorithms?  After all, Figure~\ref{fig:clumps} suggests these lattices strongly differ from random lattices, which opens the door to such a prospect.  Is it possible to specifically understand from initial principles (or perhaps even by machine learning) which parts of the lattice are not used,  and perhaps  redesign Coppersmith's method to include more useful vectors from the outset?  Finally, might the difficult issue of algebraic independence (which is needed to establish rigorously provable results) be easier to settle using these smaller sublattices?

\end{document}